\documentclass[aps,prl,twocolumn,byrevtex,superscriptaddress]{revtex4-1} 

 \usepackage{epsfig}
 \usepackage{color}
 \usepackage{amsmath}
 \usepackage{amsthm}
 \usepackage{amsfonts}
 \usepackage{amssymb}
\usepackage{graphicx}
\usepackage{epstopdf}
\usepackage[hidelinks,colorlinks=true,linkcolor=blue,citecolor=blue,urlcolor=blue]{hyperref}
\usepackage[english]{babel}

\def\be{\begin{equation}} \def\eea{\end{eqnarray}}
\def\ee{\end{equation}} \def\bea{\begin{eqnarray}}
\def\ea{\end{array}} \def\ba{\begin{array}}

\newcommand{\bel}[1]{\begin{equation}\label{#1}}

\def\zzz{{\mathchoice {\hbox{$\sf\textstyle Z\kern-0.4em Z$}}
{\hbox{$\sf\scriptstyle Z\kern-0.3em Z$}}
{\hbox{$\sf\scriptscriptstyle Z\kern-0.2em Z$}} {\hbox{$\sf\textstyle
Z\kern-0.4em Z$}}}}
\newcommand{\newG}[1]{{\textcolor{magenta}{#1}}}

\RequirePackage[normalem]{ulem}
\RequirePackage{color}
\newcommand{\new}[1]{{\textcolor{blue}{#1}}}

\begin{document}

\title{
{Turbulence Hierarchy and Multifractality in the Integer Quantum Hall Transition}
}

\author{Anderson L. R. Barbosa}
\affiliation{
Departamento de F\'{\i}sica, Universidade Federal Rural de Pernambuco, Recife-PE 52171-900, Brazil}

\author{Tiago H. V. de Lima}
\affiliation{
Departamento de F\'{\i}sica, Universidade Federal Rural de Pernambuco, Recife-PE 52171-900, Brazil}

\author{Iv\'an~R.~R.~Gonz\'alez}
\affiliation{Laborat\'orio de F\'{\i}sica Te\'orica e Computacional,
Departamento de F\'{\i}sica, Universidade Federal de Pernambuco, Recife-PE 50670-901, Brazil}
\affiliation{Unidade Acad\^{e}mica de Belo Jardim, Universidade Federal Rural de Pernambuco, Belo Jardim-PE, Brazil}

\author{Nathan L. Pessoa}
\affiliation{Laborat\'orio de F\'{\i}sica Te\'orica e Computacional,
Departamento de F\'{\i}sica, Universidade Federal de Pernambuco, Recife-PE 50670-901, Brazil}
\affiliation{
Centro de Apoio \`a Pesquisa, Universidade Federal Rural de Pernambuco, Recife-PE 52171-900, Brazil}

\author{Ant\^onio~M.~S.~Mac\^edo}
\affiliation{Laborat\'orio de F\'{\i}sica Te\'orica e Computacional,
Departamento de F\'{\i}sica, Universidade Federal de Pernambuco, Recife-PE 50670-901, Brazil}

\author{Giovani L. Vasconcelos}
\affiliation{Departamento de F\'{\i}sica, Universidade Federal do Paran\'a, Curitiba-PR 81531-980, Brazil}


\begin{abstract}

We offer a new perspective to the problem of characterizing mesoscopic fluctuations in the inter-plateau region of the integer quantum Hall transition. 
We found that longitudinal and transverse conductance fluctuations, generated by varying the external magnetic field within a microscopic model, are multifractal and lead to distributions of conductance increments (magnetoconductance) with heavy tails (intermittency) and signatures of a hierarchical structure (a cascade) in the corresponding stochastic process, akin to Kolmogorov's theory of fluid turbulence. We confirm this picture by interpreting the stochastic process of the conductance increments in the framework of H-theory, which is a continuous-time stochastic approach that incorporates the basic features of Kolmogorov's theory. The multifractal analysis of the conductance ``time series,'' combined with the H-theory formalism provides, strong support for the overall characterization of mesoscopic fluctuations in the quantum Hall transition as a multifractal stochastic phenomenon with multiscale hierarchy, intermittency, and cascade effects.
\end{abstract}
%


\maketitle




\par
{\it Introduction}.  In the 1970s Mandelbrot \cite{mandelbrot_1974} introduced multifractals as a geometrical tool to describe turbulence. Later it was discovered that the wave functions of disordered systems at the critical point of the Anderson transition exhibit strong amplitude fluctuations that can be characterized in terms of multifractal scaling  \cite{RevModPhys.80.1355,Nakayama2003,RevModPhys.67.357}. The integer quantum Hall transition (IQHT) is another important example of a quantum phenomenon in which wave function multifractalilty has been observed \cite{doi:10.1142/4335,PhysRevB.95.125414,PhysRevB.99.121301,PhysRevLett.126.056802}.  Multifractal behavior has also been observed in the magnetoconductance fluctuations in graphene-based field-effect transistors \cite{Amin2018}.
Such multifractal conductance fluctuations  have recently been characterized in mesoscopic devices subjected to a weak magnetic field in the extreme quantum regime \cite{our_paper}. 
On the other hand, turbulence-like behaviors have been identified in several physical systems besides classical fluids, such as random fibre lasers \cite{Turitsyna2013,Gonzalez2017,Gonzalez2018}, Bose-Einstein condensates \cite{PhysRevLett.103.045301}, and nonlinear optical devices \cite{Xu2015}. 
In this Letter, we report evidences of the presence of both a turbulence-like hierarchy and multifractality in the magnetoconductance fluctuations in the inter-plateau regions of the IQHT, which as far as we know has not been demonstrated before.
\par 

\new{When the system's size is smaller than the coherence length (the mesoscopic regime),  the transition between the quantum Hall plateaus presents strong conductance fluctuations which
do not behave universally \cite{PhysRevLett.59.732,PhysRevLett.69.1248,PhysRevLett.68.3757,PhysRevLett.77.4426,PhysRevB.84.245447}, in contrast with  standard mesoscopic  systems submitted to small magnetic fields where universal conductance fluctuations (UCF) are observed \cite{PhysRevLett.55.1622,1985JETPL..41..648A}.}
\new{In the traditional studies of UCF the distributions of the relevant observables tend to a Gaussian-like behavior as the average value  increases \cite{PhysRevB.49.1858,PhysRevB.51.16917,1986JETP...64.1352A,ALTSHULER1989488,Forrester_2019,PhysRevLett.88.146601,PhysRevB.81.104202,Kumar_2010}, and thus the characterization of the parametric evolution (in fictitious time) is also made via Gaussian processes \cite{RevModPhys.69.731}. 
Furthermore, the IQHT has a number of interesting questions related to subtle physical phenomena, which emerge from the interplay between quantum percolation and localization effects \cite{PhysRevB.49.4679,PhysRevB.54.R17316,PhysRevLett.82.4695,PhysRevB.66.073304,PhysRevLett.90.246802,PhysRevB.69.241305,doi:10.1143/JPSJ.77.093715,PhysRevB.72.085306,PhysRevB.80.033304,PhysRevB.81.121406,PhysRevB.102.075302,https://doi.org/10.1002/pssb.200743320} and can be traced back to the mixed nature of the dynamics inside the sample.} In classical mixed systems, regular and chaotic dynamics coexist, so that their phase spaces show a hierarchical island structure \cite{doi:10.1063/1.166481,doi:10.1063/1.166482}, which in turn invalidates standard ergodic quantum stochastic approaches, such as random matrix theory \cite{RevModPhys.69.731,Pino_2019}.
In the case of magnetoconductance fluctuations, which is the main focus of this work, one can in principle apply standard methods of stochastic physics, including field theoretical approaches \cite{efetov_1996}.
Nevertheless, it is in the framework of semiclassical techniques that real quantitative progress have been made to account for the nonergodic hierarchical structure of the phase space of mixed systems in their quantum counterparts \cite{Amin2018,our_paper,NakamuraKatsuhiro2004Qcaq,PhysRevB.54.10841,PhysRevLett.84.5504,refId0,PhysRevE.66.016211,BUDIYONO200389}. 

\par

\begin{figure}[t]
  \centering \includegraphics[width=0.5\textwidth, height=0.2\textwidth]{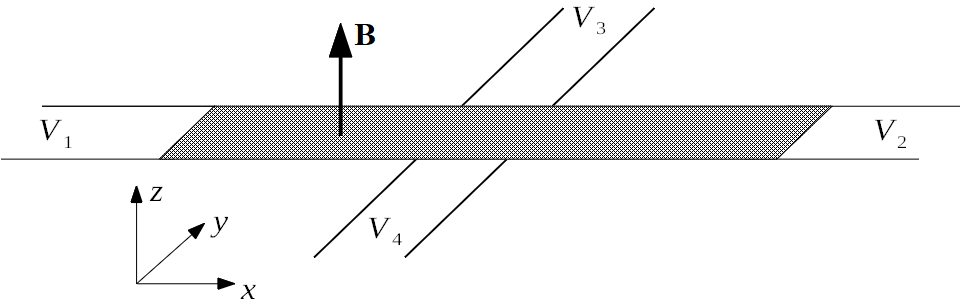}
  \caption{The integer quantum Hall setup. A disordered nanowire (shaded area) is connected to four terminals and submitted to a magnetic field ${\bf B}$ applied perpendicularly to the sample surface. The terminals are submitted to voltages $V_i$, for $i=1,...,4,$ where $V_1 > V_2, V_3, V_4$.}
  \label{fig0.eps}
\end{figure}

Here we employ a microscopic model to study a four-probe weakly disordered mesoscopic sample submitted to a perpendicular magnetic field in a regime where the IQHT is affected by strong mesoscopic fluctuations (see Fig.~\ref{fig0.eps}). 
The magnetoconductance is characterized by means of a multifractal analysis and also a multiscale hierarchical stochastic formalism, called H-theory, which has had much success in describing multifractal complex hierarchical phenomena, such as turbulence in fluids
\cite{PhysRevE.95.032315,PhysRevFluids.4.064602}, price fluctuations in stock markets
\cite{PhysRevE.95.032315} and turbulence hierarchy in a random fibre laser \cite{Gonzalez2017,Gonzalez2018,nobel}.  \new{To this end, we interpret the magnetic field as a {\it fictitious time} following previous works \cite{PhysRevLett.70.4122, PhysRevB.48.5422,PhysRevLett.70.587,PhysRevLett.70.4126,BEENAKKER199461, PhysRevLett.79.913}, which studied mesoscopic systems submitted to a small magnetic field.}
In our analysis, we implement a kind of inverse problem technique \cite{PhysRevB.102.075409} based on a running average and adapted to exhibit multifractal structure to obtain the corresponding stochastic processes as a function of external  macroscopic  parameters. The multifractal analysis gives evidence of a turbulence-like cascade process and we also find the results to be consistent with the hierarchical stochastic process of H-theory.

{\it Microscopic Model}. We consider a non-interacting two-dimensional electron gas described by the square-lattice tight-binding Hamiltonian
\begin{align}
\mathcal{H}=-t \sum_{ij}  e^{i\theta_{ij}} c_i^\dagger c_j + \sum_{i}\left(4t+\epsilon_i\right) c_i^\dagger c_i,\label{TB}
\end{align}where $c_i$ ($c_i^\dagger$) are annihilation (creation) operators and $t=\hbar^2/(2a^2m^{*})$, with $a$ and $m^{*}$ being the square lattice constant and the effective mass, respectively.
The perpendicular magnetic field ${\bf B}$ is taken into account by introducing the variables $\theta_{ij}=-e/\hbar \int_i^j {\bf A}\cdot d{\bf l}$, where ${\bf A} = \left(-B y,0,0\right)$ and $\phi=B a^2/(h/e)$ are the vector potential and the adimensional magnetic flux, respectively. The disorder is realized by an on-site electrostatic potential $\epsilon_i$ which varies randomly from site to site according to an uniform distribution in the interval $\left(-U/2,U/2\right)$, where $U$ is the disorder width. In our calculations, we used the KWANT software \cite{Groth_2014} and set $U=0.65t$ and the Fermi energy $E=1.50t$, following Ref. \cite{PhysRevB.98.155407}.

We use the Anderson model described above to study a four-probe, disordered nanowire of length $L = 310a$  and width $W = 25a$, as illustrated in Fig.~\ref{fig0.eps}, where an electronic current flows from terminal 1 to the other three terminals \cite{PhysRevLett.57.1761,datta_1995}.
The four-terminal transmission coefficients are calculated via the Landauer-B\"uttiker formula, $\mathcal{T}_{lk}=\mathbf{Tr}\left[s_{lk}^\dagger s_{lk}^{}\right]$, where $s_{lk}$ are transmission or reflection matrix blocks of the scattering matrix $\mathcal{S}=\left(s_{ij}\right)_{i,j=1,..,4}$.


\begin{figure}
    \centering
    \includegraphics[width=\linewidth]{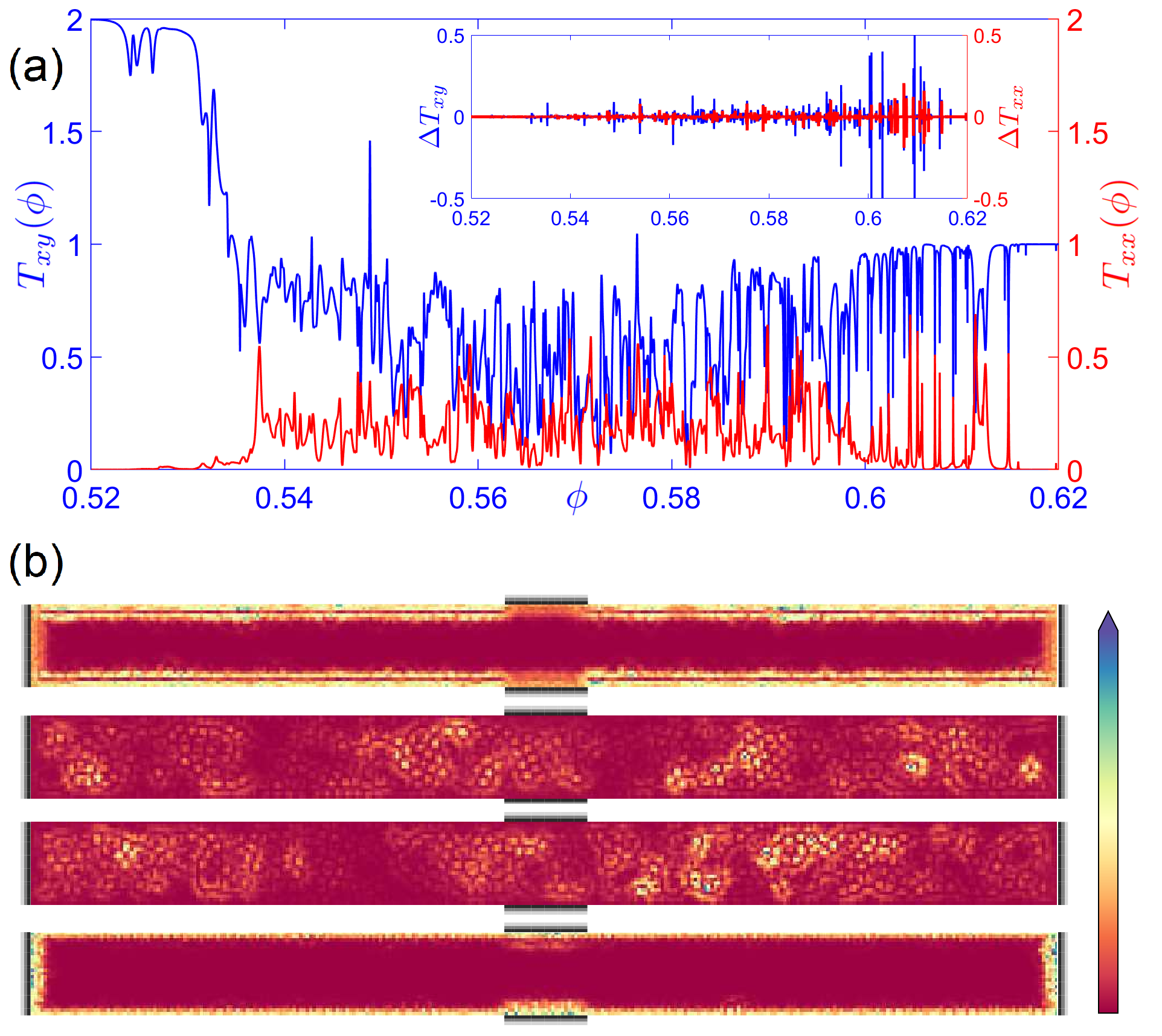}
    \caption{(a) Longitudinal, $T_{xx}$ (red), and transverse, $T_{xy}$ (blue), transmission coefficients of a disordered nanowire between the second and first Hall  plateaus as functions of $\phi$. The respective transmission increment series are shown in the inset. (b) LDOS in the transition between the second and first Landau levels for increasing values of the magnetic flux $\phi=0.50,0.55,0.56,0.64$ (from top to bottom). The probes are shown in grey. The LDOS increases as the color changes from red to blue.}
    \label{series_ldos}
\end{figure}

The transmission coefficients  $T_{xx}=\mathcal{T}_{21}$ and $ T_{xy}=\mathcal{T}_{31}+\mathcal{T}_{41}$, as functions of the magnetic flux $\phi$ (to be regarded as a {\it fictitious time}), are plotted in Fig.~\ref{series_ldos}(a) for the region between the second and first Hall plateaus. 
Note that, the coefficients satisfy the relation $T_{11}(\phi)+T_{xx}(\phi)+T_{xy}(\phi)=\mathcal{N}$, where $\mathcal{N}$ is the number of propagating wave modes in the terminals, which is tuned by the Fermi energy. 
The {\it time series} of $T_{xy}(\phi)$ and $T_{xx}(\phi)$ shown in Fig.~\ref{series_ldos} were obtained for one realization of the disorder potential with $10^4$  {\it time steps}. One sees that both longitudinal and transverse transmission coefficients fluctuate in a seemingly random fashion as the system is driven from one plateau to the next, as reported previously,  by e.g., \cite{PhysRevB.49.4679,PhysRevLett.82.4695,PhysRevLett.90.246802}. 
The corresponding series of transmission increments, $\Delta T(\phi) = T(\phi+\Delta \phi)-T(\phi)$, where $T$ stands for either $T_{xx}$ or $T_{xy}$,  are shown in the inset of  Fig.~\ref{series_ldos}(a), where one  sees that the  transmission increments fluctuate rather intermittently.

Figure~\ref{series_ldos}(b) shows colour-coded plots of the local density of states (LDOS) for increasing values of $\phi$ (from top to bottom). Note that in the Hall plateaus (top and bottom panels of Fig.~\ref{series_ldos}(b)) the LDOS is localized near the upper and lower edges of the device, as expected, since in such cases the only extended states that connect contacts are edge states. As the system is driven away from a plateau by varying $\phi$, the LDOS penetrate into the bulk and a complex spatial pattern develops (middle pannels of Fig.~\ref{series_ldos}(b)), leading to the formation of {\it coherent structures} of different sizes inside the device. This process is somewhat similar to a laminar-to-turbulence transition, in the sense that near a Hall plateau the LDOS are rather {\it laminar} (albeit restricted to the edges), whereas it becomes very irregularly distributed in space as the magnetic field is varied. In particular, one clearly sees that there is a large range of length scales in the shown patterns. Such a multiscale dynamics will be examined below from the viewpoint of both a multifractal analysis and a turbulence-like cascade model. \new{In order to perform our analysis, we conveniently chose the range of fluxes $0.545 <\phi<0.615$, which includes only transmission fluctuations values between 0 and 1, see Fig.~\ref{series_ldos}(a).} 



{\it Multifractal Analysis}.  Here we shall employ the Multifractal Detrended Fluctuation Analysis (MF-DFA) \cite{KANTELHARDT200287} 
for {\it time series} of transmission coefficients $ T(\phi)$, Fig. \ref{series_ldos}(a). For a brief description of the main steps of the MF-DFA algorithm, see the Supplemental Material (SM) \cite{supp}.
Fig.~\ref{figDFA}(a) shows the generalized Hurst exponent ${h(q)}$, which is obtained through the scaling relation of the $q$-th order fluctuation function $F_q(\tau)\sim \tau^{h(q)}$, as computed for both $T_{xy}(\phi)$ and $ T_{xx}(\phi)$ \cite{supp}. 
The strong dependence of $h(q)$ on $q$ reveals a multifractal behavior in both series. 
Fig.~\ref{figDFA}(b) shows the singularity spectra $f(\alpha)$ for $ T_{xy}(\phi)$ and $ T_{xx}(\phi)$, which both display a wide range of singularities $\Delta \alpha=\alpha_{\rm max}- \alpha_{\rm min}$, thus confirming the multifractal behavior for the magnetoconductance fluctuations in the transmission coefficients between Hall plateaus.
Furthermore, the results shown in Figs.~\ref{figDFA} (a,b)  indicate that the fictitious time series $T_{xx}(\phi)$ and $ T_{xy}(\phi)$  exhibit similar multifractal behavior, in the sense that the respective curves for $h(q)$ and $f(\alpha)$ for both series are quite close to one another.

\begin{figure}[t]
 \centering 
   \includegraphics[width=0.5\textwidth, height=4cm]{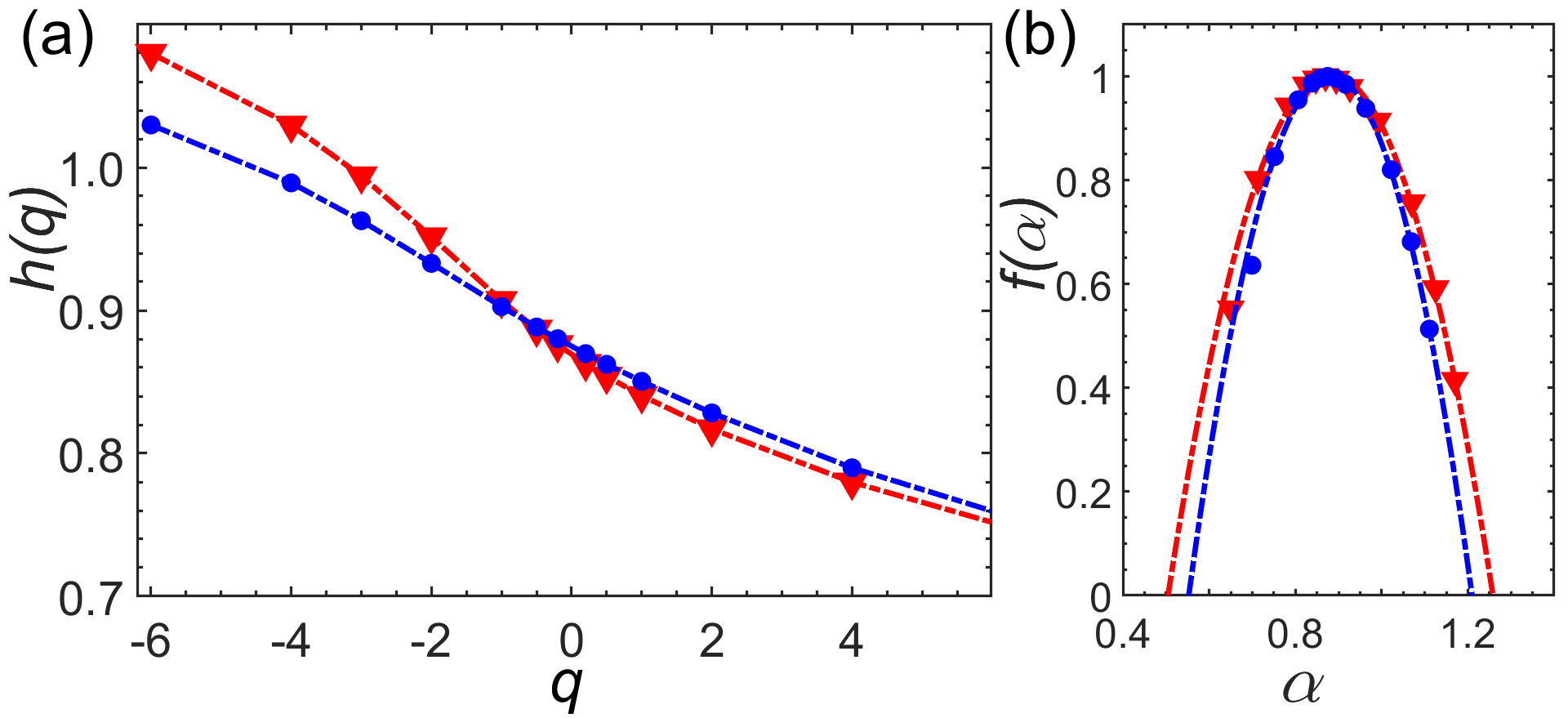}
  \caption{(a) The generalized Hurst exponent $h(q)$ 
  and (b) the  multifractal spectrum $f(\alpha)$ for the transmission coefficient series $T_{xy}(\phi)$ (blue \new{symbols}) and $T_{xx}(\phi)$ (red \new{symbols}) of the disordered nanowires. The \new{dashed} lines \new{in (a) and (b)} are \new{simple interpolations and quadratic fits of the symbols, respectively.}} 
  
  \label{figDFA}
\end{figure}

Multifractality in a time series typically has two main sources \cite{KANTELHARDT200287,Kantelhardt2011}: i) long-range correlations, which can be removed by shuffling the  series, and ii) fat-tailed distributions of the series values, in which case the multifractality cannot be removed by the shuffling procedure. To investigate the origin of multifractality in the fluctuations of the transmission coefficients  in the IQHT, 
we have computed the exponents $h(q)$ for the respective shuffled series. In all cases we found $h(q)\approx 0.5$, \new{indicating that the shuffled series are monofractal and uncorrelated, which lead us to conclude} that the multifractality in the transport coefficients in the IQHT stems mainly from correlations induced by the magnetic field variation.
Consequently, the multifractal magnetoconductance fluctuations of disordered nanowires in the IQHT points towards a complex multiscale dynamics, which we aim to investigate in detail next.



{\it H-Theory Approach.} 
Here our main objective is to find the (possibly universal) form of the distribution of the transmission increments $\Delta_\tau T(j)=T((j+\tau)\Delta s)-T(j\Delta s)$, where $j=1,...,n-\tau$, with $n$ being the number of points in our original time series $T(s)$. 
It is important to  emphasize that we seek to characterize the statistical properties of the increments, $\Delta_\tau T(j)$, rather than of the transmission coefficients $T(s)$ themselves, since time-lagged increments are more suitable quantities to investigate the scale dependence within the dynamics of conductance fluctuations in the quantum Hall problem.  
To carry out such study, we shall employ a hierarchical formalism, or H-theory for short, whose primary ideas were first considered in \cite{PhysRevE.82.047301,PhysRevE.86.050103} and later expanded in \cite{PhysRevE.95.032315,PhysRevFluids.4.064602}. 

The main mathematical aspects of the H-theory are summarized in the SM \cite{supp}.
Here it suffices to say that in this formalism the probability density function of the observed variable at short scales is written as a statistical superposition of a local quasi-equilibrium distribution, weighted by the distribution of an effective internal variable that characterizes the slowly changing background, as follows: $P(x)=\int_0^\infty P_0(x|\varepsilon) f(\varepsilon) d\varepsilon$. Here $x$ is the relevant observable, which in the IQHT problem corresponds to the transmission increments $\Delta T$;  $P_0(x|\varepsilon)$ is the local equilibrium distribution (conditioned on a fixed background), which  is assumed to have the same functional form as the large-scale distribution; and $\varepsilon$ is a random variable representing the slowly fluctuating background. For instance, in turbulence $\varepsilon$ represents the energy flux from the  adjacent larger scale eddies; whereas in the IQHT context, the background is provided by the large structures in the energy density of Fig.~\ref{series_ldos}(b), under which the `electron flow' evolves over short (fictitious) time scales. As the last and crucial ingredient of the H-theory formalism, the probability density  $f(\varepsilon)$ of the background variable is obtained explicitly from a hierarchical intermittency model in terms of special functions, namely the Meijer $G$-functions (see SM \cite{supp}):
\begin{equation}
\label{meijer1}
    f_N(\varepsilon)=
\frac{\omega}{\varepsilon_0\Gamma(\boldsymbol \beta)}    G_{ 0,N } ^{ N,0 }  \left( 
\begin{array}{c}
{-} \\ 
{ \boldsymbol\beta-{\bf 1}}
\end{array}
\bigg |\frac{\omega \varepsilon}{\varepsilon_0 }  \right),
\end{equation}
{where $N$ is the number of time scales ({\it hierarchical levels}) of the system,  $\beta_k$, $k=1,...,N,$ and $\varepsilon_0$ are parameters to be fitted to the data,  $\omega  =\prod_{k=1}^{N}\beta _k$, and we have introduced  the vector notation ${\boldsymbol\beta}\equiv (\beta_1,\dots,\beta_N)$ and 
$\Gamma({\boldsymbol\beta})  \equiv\prod_{k=1}^{N}\Gamma (\beta _k)$. In applying formula (\ref{meijer1}) to our numerical data, we shall set $\beta_k=\beta$, so that we are left with {\it only two free parameters}, namely $\beta$ and $\varepsilon_0$, for a given $N$  (see below).}

In many complex systems the large-scale increments are expected to follow a Gaussian distribution \cite{PhysRevE.95.032315}.  This is particularly true in cases where the primary quantity of interest (say, the velocity in a turbulent flow) fluctuates in a Gaussian fashion, so that the large, uncorrelated increments are expected to have the same distribution as the primary variable itself.
{Under the assumption that the large-scale distribution $P_0(x|\varepsilon)$ is a Gaussian,  the compounding integral introduced above can be performed exactly, so that the short-scale distribution $P_N(x)$ is also written in terms of $G$-functions \newG{\cite{supp}}.}
For the IQHT, however, scaling theory predicts a Gaussian distribution for the magnetoconductance fluctuations only for the bulk metallic phase \cite{doi:10.1142/4335}.
In the mesoscopic regime, i.e., when the system's size is smaller than the coherence length, there are substantial deviations from Gaussianity due to fractal fluctuations caused by interference effects \cite{PhysRevB.62.10255}.  Furthermore, 
the transmission coefficients between plateaus are constrained by the relation $T_{11}+T_{xx}+T_{xy}=\mathcal{N}$. 
Thus, in the present problem, a  Gaussian process is not expected to hold for our primary time series and, consequently, neither for the large-scale increments. We thus need a different approach to arrive at the large-scale distribution $P_0(\Delta T|\varepsilon_0)$.


It is shown in the SM \cite{supp} that {under an inverse logistic transformation, $z=2 \tanh^{-1} (\Delta T)$,} the large-scale distribution for \new{the new variable $z$} 
\new{tends to} a Gaussian function, so that the standard H-theory can be applied to $z$. Upon returning to the original variable $\Delta T$, one then obtains the following expression for the probability distribution of increments:
\begin{figure}[!]
\includegraphics[width=\linewidth]{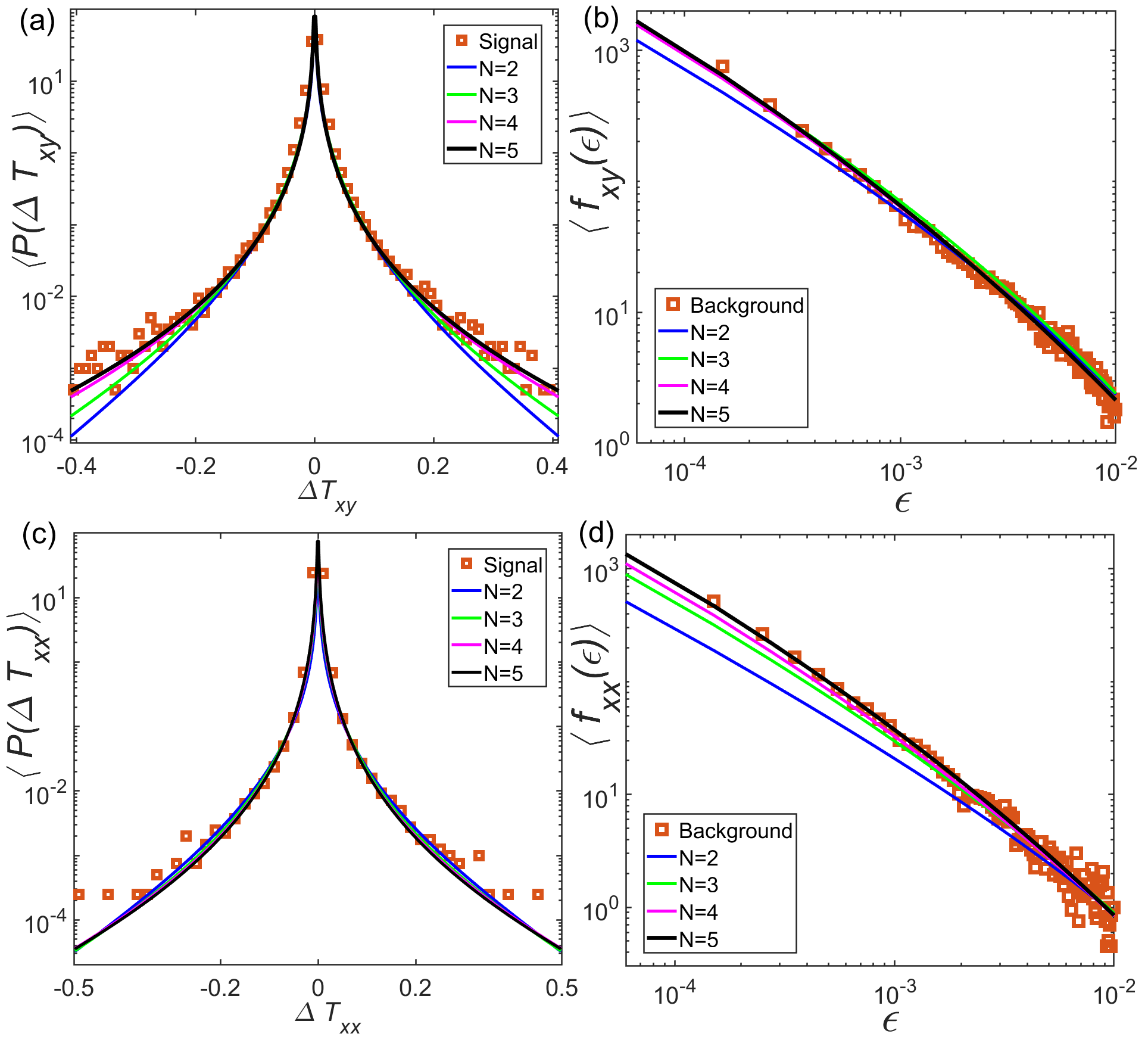}
\caption{Distribution of increments (red  squares) of (a) transversal $\Delta T_{xy}(\phi)$ and (c) longitudinal $\Delta T_{xx}(\phi)$ transmission coefficients. Continuous color lines represent the best fit of Eq. (\ref{eq:fit}) using least squares. (b,d) Histogram  (red  squares) of the variance series $\epsilon$ and model predictions (lines) fitted by Eq. (\ref{meijer1}) with the same  parameters and  color conventions as in (a) and (c).}

\label{TxiPhi}
\end{figure}

\begin{equation}
\label{eq:fit}
\begin{split}
P_N(\Delta T) = & {\frac{\sqrt{2\omega}}{\sqrt{\pi\varepsilon_0} \Gamma(\boldsymbol{\beta})}\frac{1}{{(1-(\Delta T)^2)}}}\times\\
&  G_{0,N+1}^{N+1,0}\left(
\begin{array}{c}
- \\
{ \boldsymbol\beta-{\bf 1/2}},0
\end{array}
\bigg |\frac{2\omega \left(\tanh^{-1}\Delta T\right)^2}{\varepsilon_0}\right).
%
%
\end{split}
\end{equation}
%
An important point to note is that in the H-theory approach the numerical fits are performed at the level of the background distribution given by Eq. (\ref{meijer1}). 
Only after the parameters are determined from the fit of the background distribution,  the theoretical distribution $P_N(\Delta T)$ is then plotted and compared to the empirical distribution. The agreement between theory and data in this case is  a much more stringent requirement than a direct fit to the signal distribution itself.

To proceed with the numerical analysis, we thus need to extract the background series directly from the measured data.
Once we have obtained the background series $\varepsilon_N(k)$ (see the SM \cite{supp} for details), we can then compute its histogram and fit it with the theoretical distribution $f_N(\varepsilon)$  given in Eq. (\ref{meijer1}). This allows us to infer the number of scales $N$ and  determine the best values for the parameters $\beta$ and $\varepsilon_0$. Figs.~\ref{TxiPhi}(b) and~\ref{TxiPhi}(d) show the best fits of the background distributions for $\Delta T_{xy}(\phi)$ and $\Delta T_{xx}(\phi)$, respectively, for $N=2,3,4,5$. The  plots of the theoretical distribution  $P_N(\Delta T)$, as given in Eq. (\ref{eq:fit}), are shown in Figs. \ref{TxiPhi}(a) and~\ref{TxiPhi}(c), superimposed with the respective empirical distribution for the transmission increments. Note that in both cases we observe an excellent agreement between the theoretical predictions and the empirical distributions for $N=5$, which can be understood as the number of levels in the hierarchy of length scales seen in Fig.~\ref{series_ldos}(b). This confirms that the plateau transition in the quantum Hall effect (in the mesoscopic regime)  displays a turbulence-like hierarchical  structure. 

\new{We may estimate the number $N$ of relevant length scales for a given IQH system by assuming that  the largest and smallest ``eddies'' of the electron flow in the nanowire are limited by its width $W$ and the lattice constant $a$.} \new{Supposing that these coherent structures  approximately halve in size between hierarchical levels,  we may write $W=2^N a$ \cite{PhysRevE.86.050103}. Since $W=25a$, we could expect that $N= \ln 25/\ln 2=4.6$, which is surprisingly close to $N=5$.}

{\it Conclusions}. The characterization of mesoscopic fluctuations is an important aspect of the description of phase coherent quantum transport. Our approach builds on a detailed analysis of the effective stochastic {\it time series} obtained from the longitudinal and transverse transmission increments as functions of the magnetic field in the IQHT. We have performed a multifractal analysis of the {\it time series} of transmission increments and concluded that they exhibit multifractal behavior. Besides, we also found heavy tails (intermittency) and signatures of a hierarchical structure (a cascade), similar to Kolmogorov's theory of fluid turbulence, in the transmission increments of a disordered nanowire. We interpreted the results in the framework of H-theory, which is a continuous-time stochastic approach that incorporates the basic features of Kolmogorov's theory. 
The combined descriptions---multifractal analysis and H-theory---applied here give strong evidence that the mesoscopic flutuations in the IQHT are a multiscale hierarchical turbulence-like stochastic phenomenon.

\new{Standard approaches to the IQHT apply to the thermodynamic limit, where a critical conformal filed theory can in principle be defined (for a review, see \cite{ZIRNBAUER2021168691}). However, owing to significant discrepancies between the various microscopic models, the characterization of the universality class of the IQHT is still under much debate \cite{PhysRevLett.126.076801,DRESSELHAUS2021168676}.  In the mesoscopic regime we considered here, effects such as the lattice structure may be needed to be incorporated into the effective models, so as to give rise to the emergence of hierarchical structures \cite{PhysRevB.14.2239,PhysRevLett.86.147,PhysRevLett.86.1062,PhysRevB.74.165411,Hatsuda_2016}, which are required to explain the turbulence-like features that we observed.} 
Whilst the detailed microscopic mechanism behind the cascade effect \new{in the IQHT} is  beyond the scope of the present work, we can try and interpret it in terms of an information flow between several emergent levels of descriptions in the language of non-equilibrium statistical mechanics \cite{lebellac_mortessagne_batrouni_2004}. For instance, using projection methods, the authors of Ref. \cite{RAU19961} argued that in the approach to equilibrium, information in a system may flow from slow (relevant) to fast (irrelevant) degrees of freedom, giving rise to a cascade of different Markovian levels of descriptions associated with the several stages of thermalization. In the IQHT with strong mesoscopic fluctuations, we might thus argue that akin to Kolmogorov's theory of fluid turbulence, all levels in the cascade are present at the same time giving rise to the complex pattern of information flow that is captured by the H-theory, as reported in this paper. A more detailed analysis of this argument \new{as well as a study of the possible origins of the cascade in the IQHT are} interesting topics for further research.
\new{Here we only briefly mention that the turbulence-like hierarchy  we uncovered in the magnetocondutance  fluctuations in the IQHT might be a fingerprint of  deeper underlying hierarchical structures. Possible candidates for such   structures include a) the Hofstadter butterfly energy spectrum of electrons on a 2D lattice in a perpendicular magnetic field \cite{PhysRevB.14.2239,PhysRevLett.49.405,Streda_1982, PhysRevLett.86.147}  and b) the fractal geometry of the phase space of systems with mixed dynamics \cite{our_paper}.} 



\begin{acknowledgments}
\new{After submission of the first version of this paper, we became aware of a recent work that presents experimental evidence of multifractality in the conductance fluctuations at the IQHT in a four-probe mesoscopic graphene device \cite{2112.14018}, which is in complete agreement with our theoretical results.}

This work was supported in part by the Brazilian agencies CNPq, CAPES and FACEPE.
\end{acknowledgments}

\bibliography{PRL}

\end{document}